\documentclass[nofootinbib,preprintnumbers,floatfix,twocolumn,prd,groupedaddress]{revtex4-1}

\usepackage{amsmath}
\usepackage{amssymb}
\usepackage{graphicx}
\usepackage{slashed}
\usepackage{floatrow}
\usepackage{tikz}
\usepackage{color}
\usepackage{youngtab}
\usepackage{multirow}
\usepackage{braket}
\usepackage{lipsum}
\usepackage{dsfont}

\newcommand{\degrees}{^{\circ}}
\newcommand{\R}{\mathbb{R}}
\newcommand{\F}{\mathcal{F}}

\newcommand{\origin}{\{\bf 0\}}

\begin{document}


\title{A family of repulsive neutral conductor geometries via abstract vector spaces}

\author{Julian J.~Dukes and Brian Shotwell}

\affiliation{\vspace{1mm}
Department of Physics, University of California San Diego, La Jolla, CA 92093, USA}



\begin{abstract}

Recently it was shown that it is possible for a neutral, isolated conductor to repel a point charge (or, a point dipole). Here we prove this fact using general properties of vectors and operators in an inner-product space. We find that a family of neutral, isolated conducting surface geometries, whose shape lies somewhere between a hemispherical bowl and an ovoid, will repel a point charge. In addition, we find another family of surfaces (with a different shape) that will repel a point dipole. The latter geometry can lead to Casimir repulsion.

\end{abstract}

\maketitle

\section{Introduction}\label{sec:intro}

Electromagnetism courses and textbooks usually begin with an overview of electrostatics. At the upper-division or graduate level (following, for example, Griffiths~\cite{Griffiths:1492149} or Jackson~\cite{jackson_classical_1999}, respectively), a typical unit on electrostatics will explore methods for solving Laplace's/Poisson's Equation in some region and associated uniqueness theorems. Such a discussion includes the method of images. With this method, students are expected to find, for example, the force on a point charge placed near a grounded, conducting plane, or the force on a point charge placed near a neutral, isolated sphere. For all such cases presented in these texts, the force between the point charge and a neutral conductor is attractive.\footnote{It may be ambiguous what is meant by attractive, but in cases where there is an azimuthal rotational symmetry of the conductor about some axis of symmetry, and where there is a plane normal to this axis dividing the point charge on one side and the conductor on the other, this is clear.} That is, when the point charge and conductor are held fixed in space, and when the charges on the conductor are allowed to come to electrostatic equilibrium, the resulting charge distribution on the conductor will attract the point charge.

One might naturally ask if these are examples of a more general phenomenon --- whether a point charge is \emph{always} attracted to a conductor (whether grounded, or isolated and neutral). Ref.~\cite{Levin_2011} answered this in the negative for the case of an isolated, neutral conductor, giving a particular geometry and explicitly showing the repulsion.

In this paper we arrive at another geometry where there is repulsion between an isolated, neutral conductor and a point charge. However, we do so through different means. While Ref.~\cite{Levin_2011} explicitly computed potential energy as a function of the position of the point charge along an axis of symmetry, we instead use general properties of vectors and operators on an inner-product space of scalar-valued functions defined on a two-dimensional surface.\footnote{We are not the first to cast electrostatics in this vector-space language --- see, for example, Ref.~\cite{Peter_Hahner_1999}.} Our treatment begins general, casting electrostatics in this language, but we eventually specify a particular geometry based on constraints that naturally arise in defining the force on a point charge or dipole in this setting. In addition to this method resulting in interesting repulsive geometries, we hope that this methodology can be generalized and applied to different problems altogether.

The structure of the paper is as follows: In Section~\ref{sec:vector-space} we introduce the inner-product space (which we'll also call a ``vector space"), a class of operators on the space, and some vectors in the space; this also serves to set up some notation. We also discuss how charge distributions and electric potentials can be considered vectors in this vector space, and we introduce associated operators with physical interpretation. In Section~\ref{sec:repulsive-geometries}, we consider the force on a point charge (and, afterwards, a dipole) off the surface --- how it can be cast in the language of this vector space, and how it can be repulsive in some special cases. We conclude in Section~\ref{sec:conc} with a summary and possibilities for future work.
 
\section{Vector Space; Application to Electrostatics} \label{sec:vector-space}

\subsection{Vector Space Definitions, Notation} \label{subsec:vector-space-intro}

Let $\Omega \subset \R^3 \setminus \origin$ be a bounded surface, not necessarily closed. Let $\F$ be the space of $L^2$ square-integrable, real scalar functions $\Omega \to \R$. For two such functions $f, g$, define an inner product
\begin{align}
\braket{f|g} = \int_{\vec r \in \Omega} f(\vec r)g(\vec r) \; dA \label{inner-product-def}
\end{align}
As we explore in more detail in this section, elements $\ket{\psi} \in \F$ can represent electric potentials or charge densities on $\Omega$. The bilinear scalar product (Eq.~(\ref{inner-product-def})) induces a dual space $\F^*$ with elements $\bra{\psi}$. Furthermore, $\F$ is in fact a Hilbert space, but most of the arguments presented in this paper only require general properties of an inner-product space and do not require the additional structure provided by a Hilbert space.

For every (well-enough behaved) function $O(\vec{r},\vec{s}): \Omega \times \Omega \to \R$, we can define a linear operator $O: \F \to \F$, such that
\begin{align}
O \ket{f} \equiv \ket{g}, \text{ where } g(\vec r) = \int_{\vec s \in \Omega} O(\vec r, \vec s)f(\vec s) \; dA \label{linear-operator-as-integral}
\end{align}
\noindent Every function $O(\vec{r},\vec{s})$ corresponds to an operator, but the converse is not true: there are linear operators $O: \F \to \F$ that cannot be written as integrals via Eq.~(\ref{linear-operator-as-integral}). 
We give an example in the next subsection with $O = S^{-1}$.

Then, define the following\footnote{The functions $m$, $d$, and $c$ are labeled as such to reference ``monopole," ``dipole," and ``constant," respectively. Also, Eq.~(\ref{polar-angle-def}) is defined with the negative sign since, in Section~\ref{sec:repulsive-geometries}, we will be primarily be interested in surfaces $\Omega$ that exist below the $xy$-plane. This polar angle $\theta$ is the angle that $\vec{r} \in U$ makes with the $-\hat{z}$ direction.}
 on $U \equiv \R^3 \setminus \origin$:
\begin{align}
r &= \sqrt{x^2 + y^2 + z^2} \\[5pt]
\cos\theta &= \frac{-z}{r} \label{polar-angle-def}  \\[5pt]
m(\vec r) &= \frac{1}{r} \\[5pt]
d(\vec r) &= \frac{\cos\theta}{r^2} \\[5pt]
c(\vec r) &= 1 \label{const-funct}
\end{align}
And define the following on $U \times U$ (minus the diagonal):
\begin{align}
S(\vec r_1, \vec r_2) = \frac{1}{||\vec r_1 - \vec r_2||} \hspace{0.75cm} (\vec r_1 \neq \vec r_2)  \label{S-kernel}
\end{align}
For every (well-enough behaved) function $f(\vec{r}): U \to \R$, there exists a vector $|f\rangle$, given by the restriction of the domain of $f(\vec r)$ to $\Omega$. Likewise, for functions $O(\vec{r}_1,\vec{r}_2): U \times U \to \R$, there is an operator $O$, defined via Eq.~(\ref{linear-operator-as-integral}), restricting the domain of $O(\vec r_1, \vec r_2)$ to $\Omega \times \Omega$. This defines vectors $\ket{m}$, $\ket{d}$, and $\ket{c}$, and the operator $S$.

\subsection{Charge Densities and Electric Potentials as Vectors}

$S$ has the property that its action on a charge density (via Eq.~(\ref{linear-operator-as-integral})) gives back the surface charge density's contribution to the electric potential on the surface:\footnote{This sets the potential at infinity to be zero.}
\begin{align}
S \ket{\sigma} = \ket{\phi}, \text{ where } \phi(\vec{r}) = \int_{\vec{s} \in \Omega} \frac{\sigma(\vec{s}) \, dA}{||\vec r - \vec s||} \label{charge-potential}
\end{align}
$S(\vec{r},\vec{s})$ is a Green's function of the Laplacian operator, which lets us write Eq.~(\ref{charge-potential}). However, we emphasize that $S$ is also an \emph{operator} in our vector space.

As discussed in Ref.~\cite{Peter_Hahner_1999}, $S$ is compact and self-adjoint since $S(x,y)$ is a weakly singular kernel which is real-valued and symmetric with respect to $x$ and $y$. Note $S$ being self-adjoint implies Green's reciprocity theorem:
\begin{align}
\braket{\sigma_1 | S | \sigma_2} = \braket{\sigma_1 | \phi_2} = \braket{\phi_1 | \sigma_2}
\end{align}

$S^{-1}$, if it exists, maps an electric potential to the charge distribution that produces that potential on the surface:
\begin{align}
S^{-1} \ket{\phi} = \ket{\sigma} \label{S-inv-property}
\end{align}
An explicit form of $S^{-1}$, as in the form of Eq.~(\ref{linear-operator-as-integral}), would involve the Laplacian. However, trying to do so is a little awkward given our framework: $\nabla^2 \phi(\vec{r}) = \rho(\vec{r})$ ($\vec{r} \in \R^3$) makes reference to the electric potential off the surface, and so we can't write $S^{-1}$ explicitly over the surface alone.

The operator $S$ is injective but not surjective:\footnote{One might expect $S$ to be bijective in light of existence/uniqueness theorems. Several of these theorems require continuity and/or smoothness of $\Omega$ and of the boundary value data, which we do not impose here. See, for example, Ref.~\cite{Meyers1960TheED}.}
\begin{itemize}
\item Injectivity: if there exist two distinct charge distributions that induce the same potential,
\begin{align}
S \ket{\sigma_1} = S\ket{\sigma_2} = \ket{\phi}, \hspace{0.75cm} (\sigma_1 \neq \sigma_2)
\end{align}
then their difference would be a nonzero charge distribution that produces zero electric potential everywhere on the surface. This is impossible, and so $S$ is one-to-one.
\item (non-)Surjectivity: there exist functions which may be charge distributions but cannot be potentials induced by a surface charge distribution: for example, any discontinuous function. Therefore, not every element of $\mathcal{F}$ is in the image of $S$, and $S$ is not onto.
\end{itemize}
Therefore, there exists an inverse $S^{-1}$ that is well-defined so long as we restrict the domain of $S^{-1}$ to the image of $S$. These are ``physical potentials,'' defined to be those potentials which result from a physical charge distribution on $\Omega$. We discuss the matter further and review uses of the operator $S^{-1}$ in Appendix~\ref{app:Invertibility}.

\subsection{Conducting Surfaces}

We focus now on surfaces where $\ket{\sigma}$ is not specified, but rather takes on whatever value it must to minimize the potential energy of the system. That is, we focus on conducting surfaces $\Omega$.

Suppose $\Omega$ is a neutral conducting surface (or the surface of a conducting object). Then, charges on that surface will arrange themselves so as to minimize total electric potential energy. If there are no charges outside of the surface, this is, up to a constant,
\begin{align} \label{blah}
U &= \frac{1}{2}\int_{\vec r \in \Omega}\int_{\vec s\,' \in \Omega}\frac{\sigma(\vec r) \sigma(\vec{s}\,')}{||\vec{r} - \vec{s}\,'||} \;dA \; dA' \\
&= \frac{1}{2} \braket{\sigma | S | \sigma} \label{conductor-self-energy}
\end{align}
Since this quadratic form achieves its minimum only when $\ket{\sigma} = \ket{0}$, $S$ is positive definite. This is a physical argument: the energy can be rewritten as an integral over the (non-negative) energy-density, which is non-negative and zero iff $\ket{\sigma} = \ket{0}$.

If a unit point charge is placed at the origin,\footnote{A \emph{unit} point charge at the origin normalizes the charge-distribution and electric potential on the surface $\Omega$.} then, in addition to the expression for the potential energy given in Eq.~(\ref{conductor-self-energy}), there is another term. This term, the potential energy stored in the interaction between the point charge at the origin and the surface distribution on $\Omega$, is linear in $\sigma$. Any linear functional $g: \F \to \R$ may be expressed as an inner product $\braket{g|\sigma}$; in this case, we have
\begin{align}
U = \frac{1}{2} \langle \sigma | S | \sigma \rangle + \langle m | \sigma \rangle \label{energy-conductor}
\end{align}

First, we consider conductors with no restrictions on $\ket{\sigma}$. The potential energy Eq.~(\ref{energy-conductor}) is a convex quadratic form that is minimized when $\ket{\sigma}$ satisfies
\begin{align}
S \ket{\sigma} + \ket{m} = \ket{0}
\end{align}

\noindent This is equivalent to grounding the conductor --- when energy is minimized, the electric potential due to the charge distribution on $\Omega$ and due to the point charge off the surface cancel for every point on the surface.

When the conductor is isolated and neutral, we can use 1.~the fact that the conductor is an equipotential, and 2.~the fact that the total charge is zero, to write an explicit form for $\ket{\sigma}$ in terms of $\ket{c}$, $\ket{m}$, and $S^{-1}$.

The constant potential on the surface of the conductor allows us to write
\begin{align}
S \ket{\sigma} + \ket{m} = \kappa \ket{c}, \label{const-potential-on-conductor}
\end{align}
\noindent where $\kappa$ is the constant electric potential on the conductor. Solving for $\ket{\sigma}$,
\begin{align}
\ket{\sigma} = \kappa S^{-1} \ket{c} - S^{-1} \ket{m} \label{sigma-on-conductor-part-1}
\end{align}

Total charge is another linear functional $\F \to \R$; the constraint that the total charge on the conductor is zero is equivalent to requiring $\braket{c | \sigma} = 0$. We can take the inner product of Eq.~(\ref{sigma-on-conductor-part-1}) with $\ket{c}$ to utilize this constraint and solve for $\kappa$:
\begin{align}
\braket{c | \sigma} = 0 &= \kappa \braket{c | S^{-1} | c} - \braket{c | S^{-1} | m} \\
\kappa &= \frac{\braket{c | S^{-1} | m}}{\braket{c | S^{-1} | c}}
\end{align}
Plugging this expression for $\kappa$ into Eq.~(\ref{sigma-on-conductor-part-1}) gives
\begin{align}
\ket{\sigma} = S^{-1} \ket{c} \frac{\braket{c | S^{-1} | m}}{\braket{c | S^{-1} | c}} - S^{-1} \ket{m} \label{induced-dist}
\end{align}
\noindent This gives an explicit solution for the induced charge distribution. Unfortunately, there is no explicit expression for how to compute $S^{-1}$ on the surface, so this is difficult to calculate in general.

\section{Repulsive Geometries}\label{sec:repulsive-geometries}

\subsection{Point Charge Repulsion}\label{subsec:point-charge}

A neutral, isolated conducting surface $\Omega$ that exists entirely below the $xy$-plane (i.e., negative $z$) is considered \emph{repulsive} if, when a unit point charge is placed at the origin, the charge distribution induced on the surface (from the point charge) exerts a force on that point charge whose $z$-component is positive.

As before, we use the fact that the $z$-component of the force exerted on a unit point charge at the origin is a linear functional $\F \to \R$; in this case, this is $\braket{d|\sigma}$. Substituting the expression for $\ket{\sigma} = Q\ket{m}$ from Eq.~(\ref{induced-dist}), this becomes
\begin{align}
F_z = \braket{d|Q|m}, \label{Fz-def}
\end{align}
\noindent where $Q$ is self-adjoint and defined via
\begin{align}
Q \equiv \frac{ S^{-1} \ket{c} \bra{c} S^{-1}}{\braket{c|S^{-1}|c}} - S^{-1} \label{Q-def}
\end{align}
The operator $Q$ is negative semi-definite: $\braket{x|Q|x} \leq 0$ for all $\ket{x}$. Additionally, its null space is spanned by $\ket{c}$. Proof: Since $S^{-1}$ is positive definite (on its domain of physical potentials), it defines an inner product. With respect to this inner product, $\braket{x|Q|x}$ is the norm of the projection of $\ket{x}$ onto $\ket{c}$ minus the norm of $\ket{x}$; the latter must be greater if $\ket{x} \neq k \ket{c}$, and the difference is zero iff $\ket{x} = k \ket{c}$.

This explains why the force is almost always attractive (i.e., why $F_z = \braket{d|Q|m}$ is almost always negative). The vectors $\ket{m}$ and $\ket{d}$ have a lot of overlap: the functions are each positive everywhere below the $xy$-plane, and each decreases asymptotically to zero with greater distance from the origin. It therefore makes intuitive sense that, if $Q$ maps every identical pair of vectors to a number $\leq 0$, the same would hold for most pairs that are ``close enough."

However, this is not always true --- the force is not always attractive. If, given the surface $\Omega$, there exist positive constants $k_1$ and $k_2$ such that
\begin{align}
k_1 \ket{m} + k_2 \ket{d} = k_3 \ket{c} \label{constraint}
\end{align}
($k_3 > 0$), then the surface is repulsive. The negative semi-definiteness of $Q$, along with the fact that $\ket{c}$ is in its null space, can be used to show this:\footnote{Note that, in Eq.~(\ref{mQc=0}), we took the inner product with $\ket{m}$, but using $\ket{d}$ would have worked as well.}
\begin{align}
Q \ket{k_1 m + k_2 d } &= \ket{0} \\[5pt]
\braket{ m | Q | k_1 m + k_2 d } &= 0 \label{mQc=0} \\[5pt]
\braket{m|Q|d} &= -\frac{k_1}{k_2} \braket{m|Q|m} \label{Force-Fz-is-positive}
\end{align}
\noindent $\langle m | Q | m \rangle$ is negative and $-k_1/k_2$ is also negative, and so $\braket{m|Q|d} = F_z > 0$. Therefore, $\Omega$ is repulsive.

These surfaces are parameterized by the constraint in Eq.~(\ref{constraint}). Restricting to negative $z$, they are defined by
\begin{align}
\frac{k_1}{r} + \frac{k_2 \cos\theta}{r^2} = k_3
\end{align}
We can divide by $k_3$ and rescale the constant coefficients to make them dimensionless:
\begin{align}
k_1'\frac{R}{r} + (k_2')^2\left(\frac{R}{r}\right)^2 \cos\theta = 1 \label{constraint-nondimensional}
\end{align}
We are left with a family of repulsive surfaces parameterized by $\{k_1',k_2'\}$, where $k_1'= k_1 / (k_3 R)$ and $k_2'= \sqrt{k_2 / (k_3 R^2)}$. Under the transformation $\{k_1',k_2'\} \to \{\lambda k_1',\lambda k_2'\}$, the surface is rescaled, but its shape remains the same. We can fix the scale by setting the maximum distance of the surface from the origin, achieved when $\theta = 0$, to $R$, so that the shape is controlled by a single parameter $k = k_1' = 1 - (k_2')^2$, where $k \in (0,1)$:
\begin{align}
k\frac{R}{r} + (1 - k)\left(\frac{R}{r}\right)^2 \cos\theta = 1 \label{constraint-rescaled}
\end{align}
We look at the scale-dependence of $F_z$ and introduce some dimensionless notation in Appendix~\ref{app:Scale-Dependence-Fz}; the force $F_z$ is proportional to $1/R^2$.

\begin{figure}
{\includegraphics[width=0.39\textwidth]{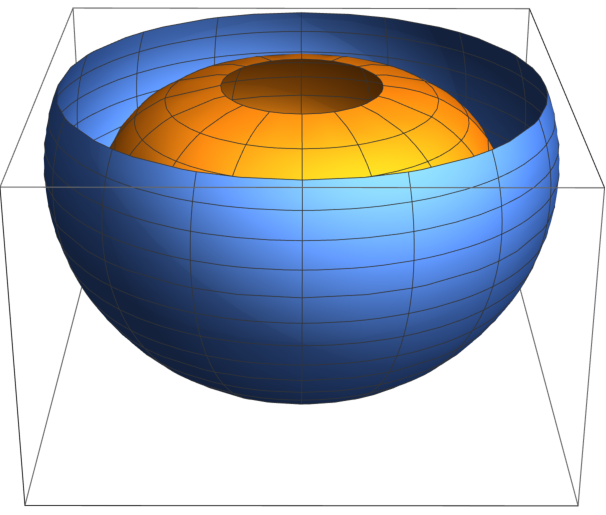}}\hspace*{0.02\textwidth}
{\includegraphics[width=0.59\textwidth]{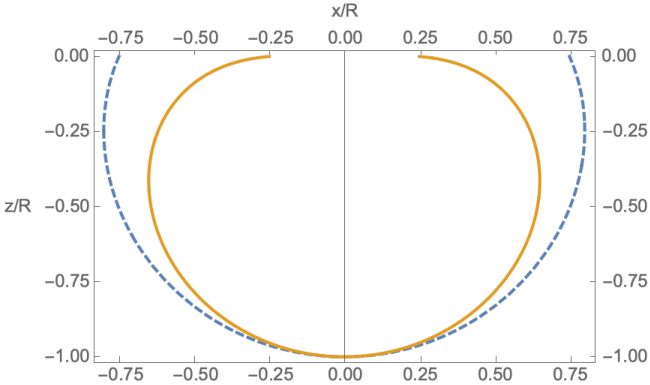}}
\caption{Plots of $\Omega_0$ (dimensionless, see App.~\ref{app:Scale-Dependence-Fz}) for point charge repulsion. FIG.~\ref{fig:point}a (left): 3D plots of surfaces restricted to below the $xy$ plane. Plotted are $k=0.75$ (blue, outer surface) and $k=0.25$ (gold, inner surface). FIG.~\ref{fig:point}b (right): 2D cross-sections of the 3D surfaces (colors the same as the 3D plot and the blue $k=0.75$ line is dashed). The limit $k \to 1$ is a hemispherical bowl.
\label{fig:point}}
\end{figure}

This phenomenon may be explained using the same intuition as Ref.~\cite{Levin_2011}: their example began with a hemisphere, then moved the hemisphere down (in the $-z$ direction) some arbitrarily small amount, causing the rim to be negatively charged and the bottom positively charged. These negative charges are closer to the point charge by an arbitrarily small amount, but their effect is attenuated by their shallow angle, so the repulsion from the positive charges dominates. These surfaces do essentially the same thing: starting with a hemisphere when $k$ is near 1, decreasing $k$ warps the rim toward the point charge, which causes the same effect. One interesting consequence that we find here is that it is not mandatory for the surface to contain all points $z<0$ that satisfy the constraint.

Our constraint gives a parameterization of the surface $r(\theta)$ that seems to imply that it must be open. However, we recognize the fact that any physical material will have a closed surface boundary. To resolve the discrepancy, we give the surface a small thickness $t \ll R$. This is similar in spirit to Ref.~\cite{Levin_2011}, where they gave their hemispherical bowl a small thickness as well.

\subsection{Dipole Repulsion, Casimir Repulsion}\label{subsec:dipole}

We now turn our attention to a \emph{dipole} placed at the origin instead of a point charge. Again, we consider a conducting surface below the $xy$ plane.\footnote{If the dipole has any finite spatial extent (which we assume is $\ll R$), we require that the conductor lie entirely below the dipole.} Equations (\ref{const-potential-on-conductor}) to (\ref{induced-dist}) apply in this scenario with the replacement $\ket{m} \to \ket{d}$; Eq.~(\ref{induced-dist}) with this replacement gives the induced charge distribution on the conductor due to the dipole placed at the origin:
\begin{align}
\ket{\sigma} = Q \ket{d}
\end{align}
\noindent ($Q$ is the same as before.) Note this assumes that the dipole moment is pointing in the negative $z$ direction.

To help in our discussion of the force on the dipole, we define one more function $q(\vec{r})$ (and associated vector $\ket{q}$) via
\begin{align}
q(\vec{r}) = \frac{3\cos^2 \theta - 1}{2r^3}
\end{align}
\noindent The $z$-component of the force on the dipole is $F_z = -F^\text{cond}_z$, where $F^\text{cond}_z$ is the $z$-component of the force on the conductor due to the dipole. By considering the $z$-component of the electric field contribution from a dipole, we can write
\begin{align}
F_z = -F^\text{cond}_z &= \int_{\vec{r} \in \Omega} \left( \frac{3\cos^2 \theta - 1}{r^3} \right) \sigma(\vec{r}) \, dA \label{Fz-dipole-integral} \\
F_z &= 2\braket{q|\sigma} = 2\braket{q|Q|d}
\end{align}

The argument is then similar to before. If, given the surface $\Omega$, there exist positive constants $k_1$ and $k_2$ such that
\begin{align}
k_1 \ket{d} + k_2 \ket{q} = k_3 \ket{c}, \label{constraint-dipole}
\end{align}
\noindent ($k_3 > 0$) then the force on the dipole is repulsive. These surfaces can be parameterized by
\begin{align}
k \left( \frac{R}{r} \right)^2 \cos\theta + (1 - k)\left(\frac{R}{r}\right)^3 \left( \frac{3\cos^2 \theta - 1}{2} \right) &= 1 \label{parameterization-dipole}
\end{align}

References \cite{Levin_2011} and \cite{Levin_2010} explain the relevance of this system to Casimir repulsion. If our dipole at the origin is replaced with a neutral, conducting needle (lying along the $z$ axis), then the instantaneous dipole-dipole forces between the needle and the conductor are repulsive, and the system exhibits the same sort of Casimir repulsion as described in these references. Through a similar analysis as that of Appendix~\ref{app:Scale-Dependence-Fz}, the force on the dipole is $\sim 1/R^4$, as expected for a Casimir force.

\begin{figure}
{\includegraphics[width=0.43\textwidth]{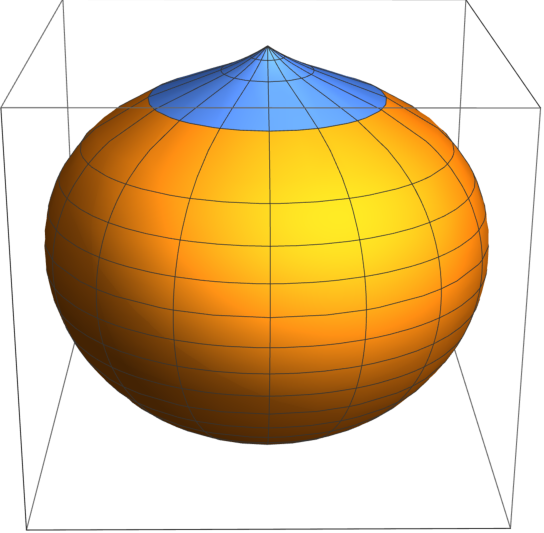}}\hspace*{0.02\textwidth}
{\includegraphics[width=0.55\textwidth]{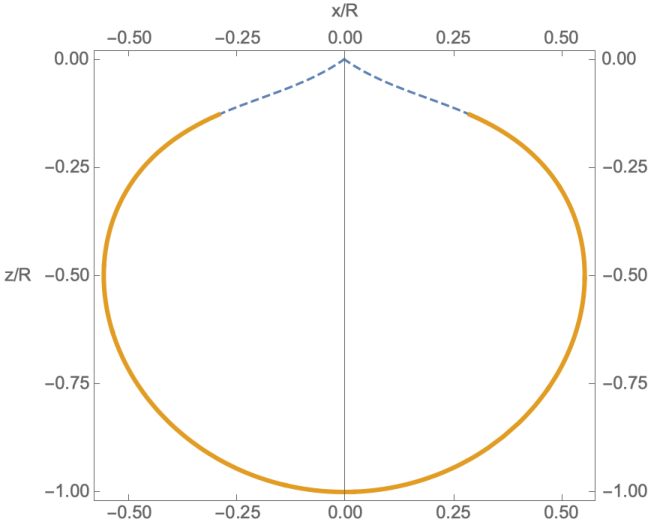}}
\caption{Plot of $\Omega_0^{k=0.75}$ (dimensionless, see App.~\ref{app:Scale-Dependence-Fz}) for point dipole repulsion. FIG.~\ref{fig:dipole-75}a (left): 3D plot of $\Omega_0^{k=0.75}$ restricted to below the $xy$ plane. The blue cap is the smaller-radius solution for those $\theta \in (\theta_\text{crit},\theta_\text{max})$ that have two solutions. (See text for details.) FIG.~\ref{fig:dipole-75}b (right): 2D cross-section.
\label{fig:dipole-75}}
\end{figure}

\begin{figure}
{\includegraphics[width=0.43\textwidth]{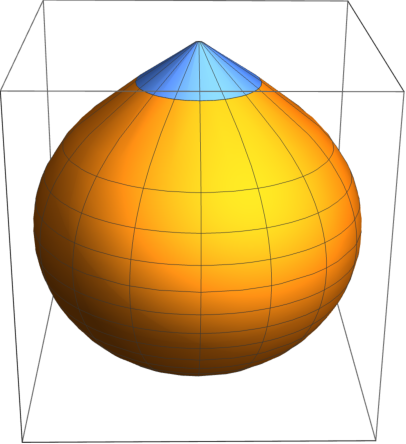}}\hspace*{0.02\textwidth}
{\includegraphics[width=0.55\textwidth]{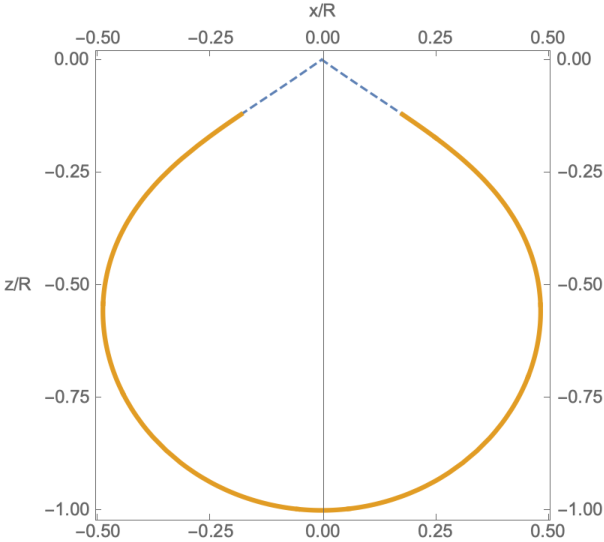}}
\caption{Similar to FIG.~\ref{fig:dipole-75}, but for $k=0.25$.
\label{fig:dipole-25}}
\end{figure}

The surfaces parameterized by Eq.~(\ref{parameterization-dipole}) have one solution for $r/R$ when $\theta \in [0,\theta_\text{crit})$, two solutions when $\theta \in (\theta_\text{crit},\theta_\text{max})$, and zero solutions when $\theta \in (\theta_\text{max}, \pi/2)$. The critical value $\theta_\text{crit} \equiv \cos^{-1} (1/\sqrt{3}) \approx 54.7\degrees$ is independent of $k$, but $\theta_\text{max}(k)$ depends on $k$.\footnote{$\theta_\text{max}(k=0.75) \approx 66.2\degrees$ and $\theta_\text{max}(k=0.25) \approx 55.8\degrees$. $\theta_\text{max}(k)$ approaches $\pi/2$ in the $k \to 1$ limit and $\theta_\text{crit}$ in the $k \to 0$ limit.}
In Figures~\ref{fig:dipole-75} and \ref{fig:dipole-25}, the ``second solution" (with smaller radius) is shown in blue; this portion of the surface is a ``cap" to the (now closed) surface, extending all the way to the origin. However, we emphasize that in Eq.~(\ref{Fz-dipole-integral}) we assumed that every part of the surface was very far away from the dipole. If the dipole has any spatial extent, then a portion of the cap of the surface must be removed, which does not change the fact that the surface is repulsive since it still satisfies the constraint.

\begin{figure}
{\includegraphics[width=\textwidth]{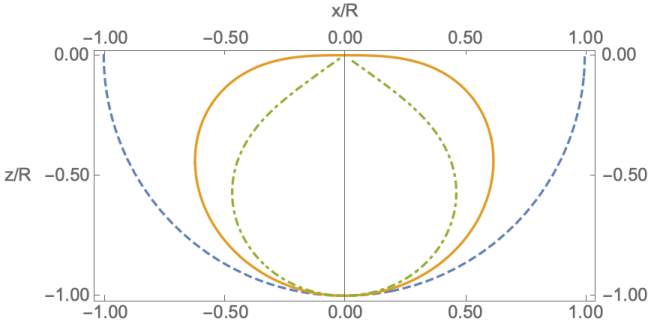}}
\caption{Level curves $m_0(\vec{u})=1$ (blue, dashed), $d_0(\vec{u})=1$ (gold, solid), and $q_0(\vec{u})=1$ (green, dot-dashed). Point charge repulsion is possible for level curves of convex combinations of the first two functions, and point dipole repulsion is possible for level curves of convex combinations of the last two.
\label{fig:all}}
\end{figure}

\section{Summary and Future Work}\label{sec:conc}
 
In this letter we have introduced an inner-product space and cast scalar-valued functions in electrostatics as vectors in this inner-product space. Using general properties of particular operators in this space, we were able to derive a class of conductive geometries that can repel a point charge. Another class of geometries can repel a point dipole, which can lead to Casimir repulsion. 

There are a few clear opportunities for future work:
\begin{itemize}
\item Either numerically or analytically, find the value of $k$ that would extremize $F_z$.
\item One might want to redefine the vector space $\F$, and possibly use a different space for charge distributions and for electric potentials, as discussed in Appendix~\ref{app:Invertibility}.
\item Instead of defining $\F$ as a single space of scalar-valued functions on $\Omega$, one could instead could have a family of surfaces that depend continuously on some parameter space, and then make a vector bundle over that parameter space, so that each vector space fiber in it varies continuously with the parameters.
\end{itemize}

\begin{acknowledgements}
The authors thank Prof.~Jeffrey M.~Rabin for very helpful comments on an earlier version of the manuscript.
\end{acknowledgements}

\appendix

\section{Invertibility of $S$}\label{app:Invertibility}

The domain of $S^{-1}$ is not the entire space, but some subspace of ``valid" potential functions, referring to functions which may represent the potential generated by a charge distribution confined to the surface. We now verify that $S^{-1}$ only acts on such vectors.

Throughout this paper, $S^{-1}$ acts only on $S \ket{\sigma}$, $\ket{c}$, $\ket{m}$, and $\ket{d}$.  The first of these is the potential on $\Omega$ produced by the charge density $\ket{\sigma}$, so this is certainly a valid physical potential. The second is a constant potential $c(\vec{r})=1$ on $\Omega$ --- this is the equilibrium situation for an isolated conductor with the appropriate amount of charge on it to produce this unit potential, absent of any external sources. The remaining two of these are potentials $\ket{\phi_\text{ext}}$ caused by an external source (a point charge or dipole at the origin). Suppose we were to ground the conductor with the external source held fixed. In that case, the induced charge distribution on $\Omega$ would exactly cancel the external potential. In other words, a charge density $\ket{\sigma_\text{ind}}$ is induced on $\Omega$ such that
\begin{align}
S \ket{\sigma_\text{ind}} = -\ket{\phi_\text{ext}}
\end{align}
This implies that the charge density $\ket{\sigma} = -\ket{\sigma_\text{ind}}$ can produce the potential $\ket{\phi_\text{ext}}$ on $\Omega$, and therefore these are ``valid" potential functions in the domain of $S^{-1}$. Furthermore, any linear combination of valid potentials is a valid potential.

In this paper, we have restricted the domain of $S^{-1}$. An alternative approach might be to find some way to treat charge distributions and potentials as two different vector spaces, where the transpose, rather than the operator $S$, maps charge distributions to the potentials they induce. We have not done this because it would prevent the map from charge to potential from having well-defined eigenvectors. Though the spectrum of $S$ is not considered here, we believe it may have interesting properties for future investigation.

\section{Scale Dependence of $F_z$}\label{app:Scale-Dependence-Fz}

Here we explore how $F_z$ for point charge repulsion depends on the length scale, $R$. Rewriting the expression for $F_z$ in Eq.~(\ref{Force-Fz-is-positive}) in terms of the rescaled constants, we have (expressing $-F_z$ in terms of both $\ket{m}$ and $\ket{d}$ to emphasize the symmetry between the two)
 \begin{align}
 -F_z &= \frac{k_1}{k_2}\braket{m|Q|m} = \frac{k_2}{k_1}\braket{d|Q|d}  \\
  &= \frac{k_1' R}{k_2'^2 R^2}\braket{m|Q|m} = \frac{k_2'^2 R^2}{k_1' R}\braket{d|Q|d}\\
  &= R^{-1}\frac{k}{1 - k}\braket{m|Q|m} = R\frac{1 - k}{k}\braket{d|Q|d}
 \end{align}
 
 We have extracted the scale-dependence of the constants $k_1$ and $k_2$, but not yet the vectors, operators, nor inner products.\footnote{For example, note $m \sim R^{-1}$ and $d \sim R^{-2}$. In addition, operators and inner products also carry some scale-dependence.} We separate the scale-dependence by defining the dimensionless functions/vectors
\begin{align}
m(\vec{r}) = \frac{1}{r} = \frac{1}{R} \frac{1}{u} &\equiv \frac{1}{R} m_0(\vec{u}) \\[5pt]
d(\vec{r}) = \frac{\cos\theta}{r^2} = \frac{1}{R^2} \frac{\cos\theta}{u^2} &\equiv \frac{1}{R^2} d_0(\vec{u})
\end{align}
$\vec{u}$ is the vector from the origin to the point on the ``unit surface" $\Omega_0$, parameterized by Eq.~(\ref{constraint-rescaled}) with $u \equiv r/R$. There is an analogous definition for $S_0$ such that\footnote{We are saying that the operator $S \sim R$. If you expected that $S \sim 1/R$ because the function $S(\vec{r}_1,\vec{r}_2)$ scales as $\sim 1/R$, note the \emph{operator} $S$ also includes an integral over the surface, and so there is an additional factor $\sim R^2$.} $S = R S_0$, and therefore $S^{-1} = S_0^{-1} / R$ and $Q = Q_0 / R$. The inner product carries some dimensions with it via the measure $dA_\Omega = R^2 dA_{\Omega_0}$; denote inner products over the unit surface via $\braket{\cdot | \cdot}_0$. Making all scale dependence explicit leaves us with
\begin{align}
F_z &= -R^{-2}\frac{k}{1 - k}\braket{m_0|Q_0|m_0}_0 = -R^{-2}\frac{1 - k}{k}\braket{d_0|Q_0|d_0}_0 \label{scale-independent-force}
\end{align}
 
As expected from dimensional analysis, the force is proportional to $1/R^2$. 

\bibliography{Conductor-Repulsion.bbl}

\end{document}